\documentclass[sigconf]{acmart}
\usepackage{balance}
\usepackage{url}
\usepackage{multirow}
\usepackage{xspace}

\definecolor{lightblue}{rgb}{0.93,0.95,1.0}
\definecolor{lightgreen}{rgb}{0.93,1.0,0.95}
\definecolor{darkgreen}{HTML}{00A64F}
\definecolor{red}{HTML}{FF0000}

\newcommand{\green}[1]{#1}
\newcommand{\red}[1]{#1}

\newcommand{\flare}{Flare\xspace}
\newcommand{\brstar}{Bert4Rec$^*$\xspace}

\copyrightyear{2025}
\acmYear{2025}
\setcopyright{cc}
\setcctype{by}
\acmConference[WWW Companion '25]{Companion Proceedings of the ACM Web Conference 2025}{April 28-May 2, 2025}{Sydney, NSW, Australia}
\acmBooktitle{Companion Proceedings of the ACM Web Conference 2025 (WWW Companion '25), April 28-May 2, 2025, Sydney, NSW, Australia}

\setcopyright{none}
\renewcommand\footnotetextcopyrightpermission[1]{}
\begin{document}

\title{FLARE: \textbf{F}using \textbf{L}anguage Models and Collaborative \textbf{A}rchitectures for \textbf{R}ecommender \textbf{E}nhancement}

\author{Liam Hebert}
\affiliation{%
    \institution{University of Waterloo}
    \city{Waterloo}
    \state{Ontario}
    \country{Canada}
}
\authornote{Equal contribution.}
\authornote{Work while done at Google Research.}

\author{Marialena Kyriakidi}
\authornotemark[1]
\affiliation{%
    \institution{Google Research}
    \city{Mountain View}
    \state{CA}
    \country{United States}
}

\author{Hubert Pham}
\authornotemark[1]
\affiliation{%
    \institution{Google Research}
    \city{Mountain View}
    \state{CA}
    \country{United States}
}

\author{Krishna Sayana}
\authornotemark[1]
\affiliation{%
    \institution{Google Research}
    \city{Mountain View}
    \state{CA}
    \country{United States}
}

\author{James Pine}
\authornotemark[1]
\affiliation{%
    \institution{Google Research}
    \city{Mountain View}
    \state{CA}
    \country{United States}
}

\author{Sukhdeep Sodhi}
\affiliation{%
    \institution{Google Research}
    \city{Mountain View}
    \state{CA}
    \country{United States}
}

\author{Ambarish Jash}
\affiliation{%
    \institution{Google Research}
    \city{Mountain View}
    \state{CA}
    \country{United States}
}

\renewcommand{\shortauthors}{Liam Hebert et al.}

\begin{abstract}
Recent proposals in recommender systems represent items with their textual description, using a large language model. They show better results on standard benchmarks compared to an item ID-only model, such as Bert4Rec. In this work, we revisit the often-used Bert4Rec baseline and show that with further tuning, Bert4Rec significantly outperforms previously reported numbers, and in some datasets, is competitive with state-of-the-art models.

With revised baselines for item ID-only models, this paper also establishes new competitive results for architectures that combine IDs and textual descriptions. We demonstrate this with Flare (\textbf{F}using \textbf{L}anguage models and collaborative \textbf{A}rchitectures for \textbf{R}ecommender \textbf{E}nhancement). Flare is a novel hybrid sequence recommender that integrates a language model with a collaborative filtering model using a Perceiver network.

Prior studies focus evaluation on datasets with limited-corpus size, but many commercially-applicable recommender systems common on the web must handle larger corpora. We evaluate Flare on a more realistic dataset with a significantly larger item vocabulary, introducing new baselines for this setting. This paper also showcases Flare's inherent ability to support critiquing, enabling users to provide feedback and refine recommendations. We leverage critiquing as an evaluation method to assess the model's language understanding and its transferability to the recommendation task.
\end{abstract}

\begin{CCSXML}
<ccs2012>
<concept>
<concept_id>10010147.10010257</concept_id>
<concept_desc>Computing methodologies~Machine learning</concept_desc>
<concept_significance>500</concept_significance>
</concept>
<concept>
<concept_id>10002951.10003317.10003347.10003350</concept_id>
<concept_desc>Information systems~Recommender systems</concept_desc>
<concept_significance>500</concept_significance>
</concept>
</ccs2012>
\end{CCSXML}

\ccsdesc[500]{Computing methodologies~Machine learning}
\ccsdesc[500]{Information systems~Recommender systems}

\keywords{Sequential and Hybrid Recommenders; Language Models}

\maketitle

\section{Introduction}
\label{sec:intro}
Modern web experiences depend on recommender systems to help users discover personalized and relevant content. A promising approach is a sequential recommender system, which incorporates the sequence of  user  interactions  to  recommend  items  grounded  in  the user’s historical preferences and temporal context \cite{Donkers2017-xy, Hidasi2016-qn, sasrec, bert4rec}. While early systems modeled sequences of items only, recent works have proposed hybrid  recommenders  that  also  model  sequential  context (e.g.  textual  descriptions,  categories,  attributes)  for  those items \cite{Hidasi2016-zt, Huang2019-pu, Zhang2019-tr}. These models suggest that the sequential item context is itself informative, with potential to improve general recommendation quality and handle cold start for new items.

With the trend in transformer-based and natural language models, more recent work has focused on recommender models that incorporate free-form text and attributes \cite{recformer, Li2024-cx, s3rec, carca, u-bert, userllm, tiger}. Those papers show that using text tends to improve performance over item ID-only baselines, e.g. Bert4Rec \cite{bert4rec}, which is widely used as a baseline to justify new approaches. However, we discovered that when thoroughly tuned, Bert4Rec surpasses the performance of text-based recommenders, on certain datasets. On datasets where Bert4Rec underperforms, our tuned model consistently outperforms the previously reported Bert4Rec baselines often used to evaluate novel recommenders. These findings suggest that ID-only models are more competitive than assumed.

With revised ID-only baselines, we investigate the performance of hybrid models that combine IDs and text. This paper presents \flare, a simple test-bench hybrid recommender that extends Bert4\-Rec by incorporating a modern text encoder using a Perceiver \cite{perceiver} network. This integration enables the synthesis of text-based item context and item embeddings, potentially providing a richer representation for recommendations. \flare outperforms state-of-the-art models on some datasets, suggesting that the combination of IDs and text unlocks additional gains.

To the best of our knowledge, the effectiveness of recommenders in the literature is mostly evaluated on small datasets, specifically ones with a limited number of items. For instance, many models use subsets of the Amazon Product Reviews dataset containing no more than around 40k, and often significantly fewer. In contrast, commercially-applicable recommenders must handle much larger item spaces. This gap in scale invites questions on whether contemporary approaches that combine IDs and rich metadata are still performant and feasible in a more realistic setting. We answer this with new baseline results for the Clothing, Shoes and Jewelry dataset from Amazon Product Reviews, which involves a larger vocabulary (376k). We observe similar performance trends as in smaller datasets. \flare also outperforms other contemporary models when fine-tuned and evaluated on this dataset.

Finally, while it is natural to use standard ranking metrics to evaluate hybrid recommenders, this paper proposes critiquing \cite{Chen2012-dv, Gao2021-xo, Jannach2021-bp} as an additional methodology. For   recommenders trained with textual context, we believe that evaluating recommendations with text critiques is a natural way to measure how well the model assimilates item and textual information. To evaluate models with critiquing, we simulate the user providing input for the desired recommendations based on recent recommended items. We present an evaluation methodology for applying critiquing using the Amazon Reviews dataset and establish baselines.

\begin{figure*}
    \centering
    \includegraphics[width=0.8\linewidth]{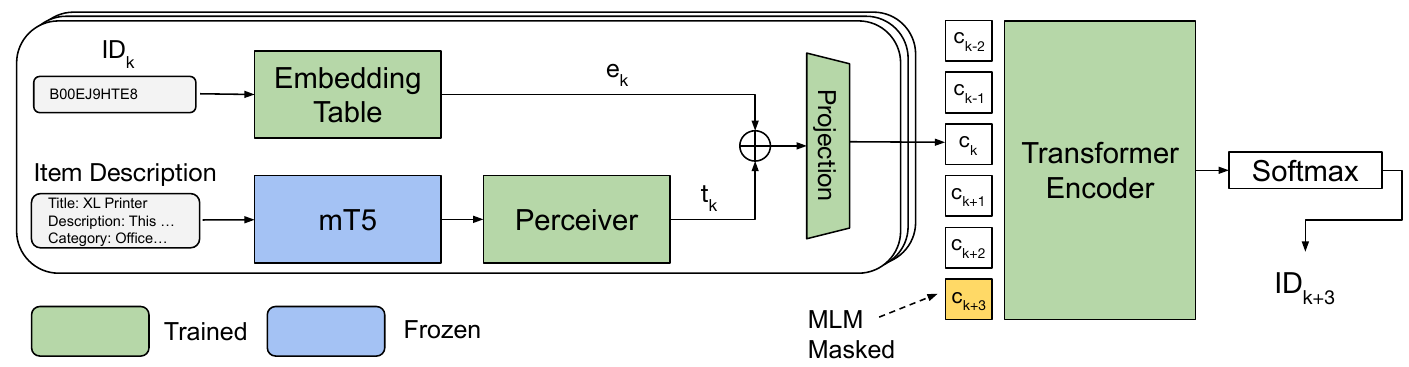}
    \caption{Sequential recommender model architecture with text encoder augmentation. Text tokens are encoded with a frozen (blue) language model (mT5), and the associated embeddings are resampled with the Perceiver. Parameters in green boxes are learned using masked language modeling (MLM).}
    \label{fig:arch}
\end{figure*}

Our contributions include:
\begin{itemize}
\item Updated Bert4Rec baselines with thorough hyperparameter tuning, where they consistently outperform previously reported Bert4Rec runs, and in some datasets, outperform state-of-the-art models. They are valuable for reinterpreting conclusions from prior work that use Bert4Rec as a baseline, e.g. that the performance gains of adding text may not be as large as previously thought. 

\item \flare, a hybrid recommender model using a language model to encode text and a Perceiver network to produce contextual embeddings (Figure~\ref{fig:arch}). These are then combined with ID embeddings for a richer representation. We show that \flare obtains competitive performance on standard datasets, re-affirming the value of text metadata.

\item New baselines for a larger vocabulary dataset: the Clothing, Shoes and Jewelry dataset from Amazon Product Reviews (376k), $\sim$10x larger than commonly-used datasets. We establish robust baselines on an updated Bert4Rec, fine-tuned Recformer~\cite{recformer}, and 3 model sizes of \flare, all of which serve as benchmarks for future work on large item vocabularies.

\item Critiquing as a promising methodology to assess performance gains obtained from adding text metadata. We present a new critiquing-based metric and associated baselines.
\end{itemize}

This paper is organized as follows. The next section overviews related work. Section~\ref{sec:method} describes \flare's architecture, after which Section~\ref{sec:eval} presents our experiments and evaluation benchmarks. Finally, Section~\ref{sec:conclusion} outlines future work and concludes.
\section{Related Work}

Work in sequence modeling for recommendations have traditionally focused on RNNs \cite{Hidasi2016-qn, Donkers2017-xy}, and more recently, using transformers. For example, SASRec \cite{sasrec} proposes using a transformer-based approach to model item sequences for recommendations, using causal self-attention. Bert4Rec \cite{bert4rec} extends SASRec with BERT-style \cite{bert} masked language-model training for item sequences. These seminal models serve as popular baselines in subsequent work.

To improve general recommendation quality, follow-on efforts investigate hybrid models, which augment the sequence of item IDs with a corresponding sequence of item context (e.g. textual descriptions, categories, attributes). For example, HybridBERT4Rec \cite{hybridbert4rec} extends Bert4Rec with a separate tower that models item context, using the set of users that consume that item as input. S$^3$Rec \cite{s3rec} proposes self-supervised pre-training tasks to align items and attributes by utilizing mutual information maximization. CaFe \cite{li2022coarsetofine} explicitly models coarse-grained, or intent-based, sequences along with dense fine-grained sequences to better capture short-term vs long-term preferences. CARCA \cite{carca} combines item sequences, associated context, and item attributes connected via cross-attention to better correlate between older and more recent items in the user profile. U-BERT \cite{u-bert} trains user and content encoders during self-supervised pre-training with content-rich signals, followed by fine-tuning to focus on content-deficient domains.

Some efforts remove dependence on an explicit item ID representation, relying solely on a contextual representation. For example, Recformer \cite{recformer} proposes an item representation comprising learned embeddings of text-based key-value pairs. TIGER \cite{tiger} investigates a semantic ID representation for items, along with a T5-style \cite{T5} training and generative approach for retrieval. UniSRec~\cite{Hou2022-xt} learns transferable item representations based on item description, along with sequence representations with contrastive tasks. UniSRec can also incorporate IDs in a fine-tuning step.

As described in Section~\ref{sec:method}, \flare shares similarities with some of the models above. It learns embeddings to represent each item, uses a pre-trained text encoder to encode item metadata, and models item sequence with a stack of transformer layers. \flare differs in that it does not require additional pre-training (aside from the ``off-the-shelf'' text encoder) nor fine-tuning.

Recent efforts explore incorporating collaborative filtering signals directly into a large-language model (LLM), motivated by the intuition that the language semantics encoded in the LLM promote item context interpretation. P5 \cite{p5rec} represents all user--item signals, along with associated context, as natural language sequences that are input to an encoder/decoder model. LC-Rec \cite{lc-rec} uses vector quantization for ID representation and LLM pre-training tasks to align items and language. CLLM4Rec \cite{zhu2023collaborative} adds new user and item tokens to the LLM, along with a training head to model the collaborative filtering signal. Hua et al. \cite{Hua_2023} study methods to effectively add item IDs to the LLM token vocabulary. User-LLM \cite{userllm} encodes user history into user embeddings and incorporates them into an LLM via cross-attention. Harte et al. \cite{Harte2023-gq} experiment with extracting item embeddings from a pretrained LLM to make recommendations directly, as well as to initialize Bert4Rec's item embeddings. CALRec \cite{Li2024-cx} proposes a two-stage LLM finetuning approach using a combination of contrastive and language modeling losses. Unlike the above, \flare is focused on using encoders and does not explore the use of a language decoder to make recommendations.

We propose critiquing (surveyed in \cite{Chen2012-dv, Gao2021-xo, Jannach2021-bp}) as a new task for evaluating hybrid models. Early critiquing systems enable the user to provide feedback on recommended items, typically at the feature-level, through (e.g.) suggested pre-designed tweaks \cite{Burke1996-rv} or language-driven dialog systems \cite{Shimazu2002-uq, Thompson2004-jj, Grasch2013-kw}. Recent work \cite{Wu2019-qx, Luo2020-uw, Luo2020-qh} explores systems that jointly learn latent representations for users, items, and key-phrases, the latter of which serves as critiques that can influence the recommendations through operations in the latent space. Others explore the interpretation of attributes \cite{Nema2021-ad, Balog2021-cj} to support critiquing. We view hybrid models that associate item sequences with the associated textual context signals as natural extensions to the recent critiquable recommenders. We believe hybrid recommenders offer a promising approach for building critiquing systems.

\begin{figure}
    \centering
    \includegraphics[width=0.9\linewidth]{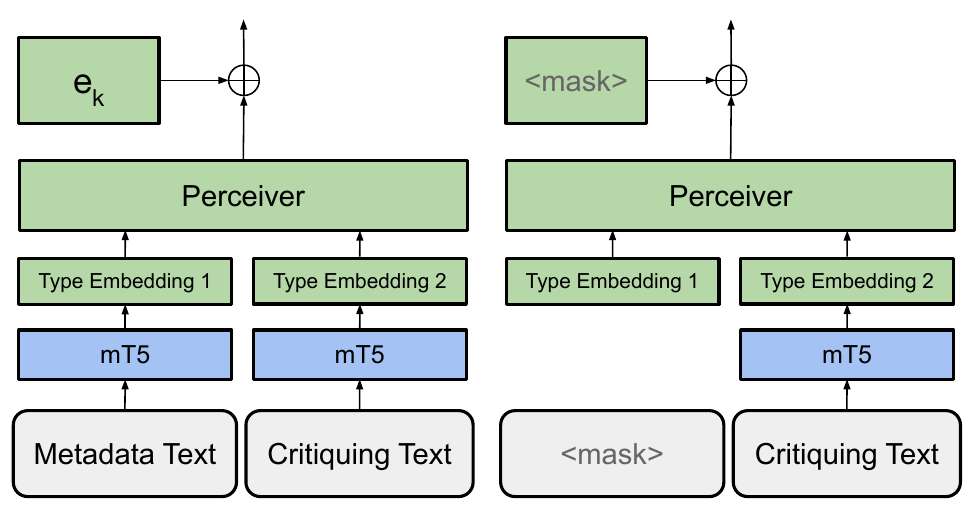}
    \caption{Flare adapted to incorporate recommendation critiques. Critiques are textual descriptions, incorporated as a separate modality to the Perceiver, to steer the model's predictions.}
    \label{fig:perceiver}
\end{figure}

\section{Method}
\label{sec:method}
\subsection{Task Definition}
We formulate the recommendation task as follows. A dataset consists of user sequences, with each sequence $U = \{i_1, i_2, ..., i_n\}$ containing items $i \in \mathcal{I}$. Each item $i$ is a tuple $(\mathrm{ID}, T)$, where $\mathrm{ID}$ is the unique identifier of the item, and $T = \{t^1, t^2, ..., t^m\}$ is a natural language description of the item consisting of text tokens $t$. Each description $T$ comprises textual content (e.g. item title, category, brand, price, etc). The task is to predict $i_{n+1}$ given $U$.

\subsection{\flare Model Architecture}

Flare augments the Bert4Rec architecture \cite{bert4rec} with item context features (Figure \ref{fig:arch}). Sequential recommendation models like Bert4Rec use masked language modeling (MLM) with a cloze learning task, similar to BERT's approach for language tokens. During training, a random subset of items in the input sequence is replaced with a special mask token $\langle mask \rangle$. The model's objective is to predict these masked items based on their surrounding context.
For example, an input sequence of items $[ID_1,ID_2,ID_3,ID_4,ID_5]$ could be transformed through random masking into $[ID_1$, $\langle mask \rangle_1$, $ID_3$, $\langle mask \rangle_2$, $ID_5]$. The labels for this masked sequence would then be $\langle mask \rangle_1=ID_2$, $\langle mask\rangle_2=ID_4$.

Following the item embedding lookup, the inputs to the initial transformer layer are represented as:
\begin{equation*}
[e_1, e_{\langle mask \rangle},e_3,e_{\langle mask \rangle},e_5]
\end{equation*}
where $e_i$ represents the item embedding for item $ID_i$, and $e_{\langle mask \rangle}$ denotes the mask embedding.

The transformer layers process the masked sequence, and the outputs corresponding to the masked positions are fed into a softmax function over the item vocabulary. The model's loss is calculated as the average negative log-likelihood over the masked positions:
\begin{equation}
\mathcal{L}_{MLM}=\frac{1}{|U^m|}\sum_{ID_m \in U^m} P(ID_m=ID_m^*|U')
\end{equation}
where $U^m$ is the set of masked items,  $ID_m^*$ is the actual item corresponding to the masked item $ID_m$,  $U'$ is the input sequence with masking and $P(\cdot)$ represents the output of the softmax layer given the masked input.

To incorporate item contextual information, we augment the item embeddings in the masked sequence as
\begin{equation*}
[c_1,e_{\langle mask \rangle},c_3,e_{\langle mask \rangle},c_5]
\end{equation*}
where $c_i$ is derived from a combination of the item embedding $e_i$ and the item context embedding $t_i$ representing item metadata.
$t_i$ is generated using a pre-trained and frozen LLM encoder and a perceiver \cite{perceiver} to encode the LLM output embeddings into a fixed length. The output of the perceiver is then linearly projected to match the item embedding dimensions and then vector summed with the item embeddings. This allows the model to learn masked item prediction with \textit{text as side information}, which we show improves the performance on recommendation tasks. In our experiments, we use the mT5 Base model \cite{mt5} to encode text. 

The use of a perceiver network provides certain key advantages in this context. It encodes a variable length input sequence of embeddings to a fixed length output sequence of embeddings. By concatenating (in the sequence dimension) different modality types at its input, the network can learn a joint encoding. Thus the perceiver offers a natural method of joining multiple modalities of text data, as described later for critiquing.

\subsection{ID-Text Contrastive Learning}
 We implement an auxiliary contrastive loss to align item ID embeddings with item context embeddings. This approach is motivated by the observation that the less frequent, or ``tail'' items, which have limited representation in training data, can rely instead on their contextual representations. Conversely, frequently appearing items can benefit from collaborative filtering signals that may be learned from training a sequential recommender. Aligning ID and context embeddings places both the collaborative filtering (CF) and contextual signals into the same latent space, allowing context to potentially fill in for gaps in CF. More precisely, we employ the InfoNCE loss function \cite{infoNCE} as detailed below:
\begin{equation}
        \mathcal{L}_c = - \frac{1}{N}\sum^N_{i=1}
        \log \frac{\exp(\frac{e_i \cdot t_i}{\tau} - m)}{\exp(\frac{e_i \cdot t_i} {\tau} - m) + \sum^N_{j = 1, j \neq i} \exp(\frac{e_i \cdot t_j}{\tau})}
\end{equation}
where $\tau$ is the softmax temperature, and we also include a softmax additive margin term $m$ to each positive pair to increase the separation between positive and negative pairs, as proposed by \cite{yang2019improving}.

\subsection{Critiquing}
\label{sec:method_critique}

A core component of hybrid recommenders is their ability to effectively utilize both textual content semantics and collaborative filtering information. However, bridging the gap between these two modalities poses a challenge, potentially leading to a situation where the downstream sequential recommender fails to fully exploit the information available in each.

We propose a novel masking and evaluation paradigm using critiquing to assess whether hybrid recommenders can appropriately leverage both modalities. This consists of adding a critique in the form of a textual item category $S$ to each item $i$ in the user's sequence $U$, providing a broad hint to direct the model to the next recommended item. That is, each item $i$ in the user sequence $U$ becomes $i = (ID, T, S)$, where $S$ is the critique string. For masked locations during training and inference, we only mask the $ID$ and item text $T$, leaving $S$ unmasked, which the model can utilize as a hint towards the target item (i.e., $i = ([\langle mask \rangle, \langle mask \rangle, S)$).

The key to this task is that hints are textual, therefore requiring recommenders to adapt to a text modality by relating the critique to the other signals in the user's sequence.
Note that this task formulation has applications to many practical uses of recommender systems, such as refining recommendations based on the user's natural language query text or the webpage subsection the user is currently navigating. In essence, critiquing strings can help steer a model's predictions to incorporate the user's current context.

We incorporate critiquing text in \flare as shown in Figure \ref{fig:perceiver}. Critique text is a distinct type of input modality, separate from the item contextual text. This is done by separately encoding each modality, with distinct type embeddings, and then using the Perceiver network to merge both text modalities together. Given an item $i_k = (ID_k, T_k, S_k)$ consisting of an $ID_k$, natural language description $T_k$ and critique $S_k$, we compute the modified item contextual representation $t_{k}$ and augmented item representation $c_k$ with critiquing text as
\begin{align}
    T'_k &= LLM(T_k) + \theta_T \\
    S'_k &= LLM(S_k) + \theta_S \\
    t_k &= Linear(Perceiver([S^{'}_{k} ; T^{'}_{k}])) \\
    c_k &= t_k + e_k
\end{align}
where $\theta_T$ and $\theta_C$ are type embeddings to denote text and critiquing as distinct types, ``$;$'' denotes concatenation, ``$+$'' denotes vector addition, and $e_k$ is the embedding for item $ID_k$. When masking items for training or during inference, we replace $T'_k$ and $e_k$ at the desired positions with $\langle mask \rangle$ embeddings, allowing the model to utilize $S'$ as a hint. By treating critique text as a separate input modality, we hypothesize that the model can learn the connection between critique text and item context.
Note that critiquing is applied with Flare through an optional additional input to the Perceiver model, which can be omitted.

\subsection{Training}
\label{sec:training}
We train \flare using the MLM loss by masking items in the user's sequence with 15\% probability, ensuring that at least one item is masked. For the total loss, we use a weighted sum of MLM loss and the ID-Text contrastive loss:
\begin{equation}
    \mathcal{L}_{total} = \alpha \mathcal{L}_{MLM} + (1 - \alpha) \mathcal{L}_{c}
\end{equation}
controlled by the hyper-parameter $\alpha$, which we found $\alpha=0.5$ yielded optimal results. For the contrastive loss, we use a softmax temperature $\tau = 0.3$ and softmax additive margin term $m = 0.2$.

\section{Experimental Results}
\label{sec:eval}

This section investigates the following research questions:
\begin{itemize}
    \item \textbf{RQ1}:
     Do the previously reported Bert4Rec baseline models have potential for further performance gains through tuning?
    \item \textbf{RQ2}: Can hybrid models like \flare perform well on larger item count datasets?
    \item \textbf{RQ3}: Can \flare relate textual critique commands to sequential user signals?
    \item \textbf{RQ4}: Can \flare adapt to critique commands that are out of sequence?
\end{itemize}

\subsection{Datasets}

\begin{table}
    \centering
    \caption{Statistics for the Recformer- and CALRec-processed Amazon Product Reviews dataset. I/U is the average number of items per user, or average sequence length. Note that the Clothing dataset preprocessing follows that of Recformer, except that it filters out long sequences rather than trimming them (Section~\ref{sec:clothing_dataset}).}
    \begin{tabular}{l r r r r r r}
        \toprule
        & \multicolumn{3}{c}{Recformer} & \multicolumn{3}{c}{CALRec} \\
        \cmidrule(r){2-4} \cmidrule(l){5-7}
        Dataset & \# Users & \# Items & I/U & \# Users & \# Items & I/U  \\
        % \midrule
        \cmidrule(r){2-4} \cmidrule(l){5-7}
        
        Games & 55,132 & 17,389 & 9.00 & 50,940 & 17,383 & 8.97 \\
        Office & 101,357 & 27,932 & 7.88 & 44,736 & 27,482 & 7.87 \\
        Scientific & 11,011 & 5,327 & 6.98 & 9,461 & 5,282 & 7.04 \\
        Music & 27,525 & 10,611 & 8.40 & 25,577 & 10,599 & 8.39 \\
        Arts & 56,110 & 22,855 & 8.77 & 47,197 & 22,828 & 8.72 \\
        Pets & 47,569 & 37,970 & 8.84 & 43,135 & 37,712 & 8.82 \\
        \midrule
        Clothing & 1,219,592 & 376,843 & 9.20 \\
        \bottomrule
    \end{tabular}
    \label{tab:datasets}
\end{table}

We use datasets from Amazon Product Reviews \citep{ni-etal} to evaluate our model. The Amazon Product Reviews dataset contains 233.1 million reviews from May 1996 to October 2018, covering 29 categories ranging from Office Supplies to Video Games. Each review includes item features (title, description, category, price) and review features (text, timestamp, score, summary). User sequences are created by grouping and sorting reviews by timestamp, enabling a sequential recommendation task to predict the next item the user reviews.

We benchmark against Recformer~\cite{recformer} and CALRec~\cite{Li2024-cx}, which report the current state-of-the-art results. To meaningfully compare our results to each of those projects, we use the same Amazon Product Reviews dataset preprocessing procedure that each of those projects use. Where available, we either downloaded and trained with the relevant preprocessed datasets or ran the project's pipeline to preprocess the raw dataset. Dataset statistics are in Table~\ref{tab:datasets}.

\subsubsection{Recformer-Processed Dataset}
\label{sec:recformer_dataset}
We use the six dataset splits provided in Recformer:\footnote{We observed that the statistics from the downloaded Recformer datasets differ from the statistics reported in the Recformer paper, most notably the Games dataset.} ``Video Games'', ``Office Supplies'', ``Industrial and Scientific'', ``Arts, Crafts and Sewing'', ``Pet Supplies'' and ``Musical Instruments.''

\subsubsection{Clothing Dataset}
\label{sec:clothing_dataset}
In addition to the six small datasets above, we also evaluate on one medium-sized dataset, ``Clothing, Shoes and Jewelry,'' which contains 378k items. The preprocessing procedure is the same as Recformer, but rather than trim, we filter all user sequences with more than 50 items and items with a missing title. Notably, this dataset is approximately 9x larger than the largest small dataset.

\subsubsection{CALRec-Processed Dataset}
\label{sec:calrec_dataset}
CALRec preprocessing notably performs data-deduplication on user sequences. We follow CALRec's deduplication methodology and remove consecutive items by the same user that have the same item ID (ASIN number), review text, rating, and timestamp. The motivation for deduplication is to mitigate trivial recommendation strategies like always recommending the most recent item.

\subsection{Experimental Setup}
\subsubsection{Baselines}
We compare \flare against three classes of state-of-the-art sequential recommenders.

First are ID only models, like Bert4Rec \cite{bert4rec}, which utilize a transformer-based approach with BERT-style \cite{bert} masked language-model training for item sequences. We show updated baselines  after extensive tuning, which we call \textbf{\brstar}. We emphasize that \brstar is Bert4Rec, with our hyperparameter tuning.

Second are text-only models, which use a text-based representation of user history and items without an explicit item ID representation. We focus benchmarking against Recformer~\cite{recformer} and CALRec~\cite{Li2024-cx}, which report the current state-of-the-art results.

Finally, we compare against hybrid Text + ID models, specifically (a) S$^3$Rec \cite{s3rec}, which utilizes self-supervised pre-training tasks to align items and attributes through mutual information maximization; and (b) UniSRec \cite{Hou2022-xt}, which relies on text-based item representations but can incorporate IDs in a fine-tuning step.
 
One notable difference between \flare and the prior models Recformer and CALRec is that \flare trains a separate model for each dataset. In contrast, Recformer and CALRec propose a two-stage training approach, in which a single model is pre-trained on multiple datasets, followed by a fine-tuning stage on reported evaluation datasets. We view this as a difference in model training approach.

\subsubsection{Evaluation Settings}
\label{sec:eval_settings}
Following prior work \cite{recformer,lc-rec,bert4rec}, we adopt the leave-one-out strategy for evaluation. We reserve each user's penultimate and last items for validation and test, respectively, using the rest of the sequence for training.

Since training sequences can be of varying length (Table \ref{tab:datasets}), we pack multiple training sequences into a single input example such that each packed example is as close to the same size as possible. Then, during training, we restrict attention to just the items within the same sequence. This maximizes the throughput of our hardware and avoids wasteful computation due to excessive padding tokens. As a result, the effective batch size during training can vary from step to step, depending on the length of the sequences.

For our experiments, we use the Adam optimizer with $\beta_1=0.9$ and $\beta_2=0.99$, following standard practice \cite{Reddi2018-ne}. We also performed a hyperparameter sweep on: the number of the transformer's layers and heads, model and hidden dimensions, learning rate, batch size, total step count, weight decay, usage of contrastive loss, number of Perceiver layers, and number of heads and latents for the Perceiver network. Appendix~\ref{sec:appendix_hyperparameters} summarizes the best model configurations for each dataset.

\begin{table*}
    \setlength{\tabcolsep}{1.5pt}
    \centering
    \caption{\flare performance against prior baselines, for both Recformer-processed \cite{recformer} and CALRec-processed \cite{Li2024-cx} datasets. Baseline results are quoted from those papers. For CALRec, we cite the optimistic (better) numbers above. \brstar is our tuned Bert4Rec, which outperforms previously reported Bert4Rec baselines. The \%SotA column denotes the performance difference between \flare and the best prior model reported, and \%ID is the difference between \flare and \brstar. The best result is \textbf{bolded}; the second best is \underline{underlined}. Note: Recformer-Processed UniSRec is a text-only model, whereas CALRec-Processed UniSRec$^{\dagger}$ is an ID + Text model. \flare achieves competitive performance.}
    
    \resizebox{\textwidth}{!}{%
    \begin{tabular}{p{1.2cm} l cccccccc ccccccccc}
        \toprule
        & & \multicolumn{8}{c}{Recformer-Processed Dataset} & \multicolumn{9}{c}{CALRec-Processed Dataset} \\

        \cmidrule(r){3-10} \cmidrule(l){11-19}
        Dataset  & Metrics & Bert4Rec & UniSRec & Recformer & S$^3$Rec & \brstar & \flare & \%SotA & \%ID & 
            Bert4Rec & Recformer & S$^3$Rec & UniSRec$^{\dagger}$ & CALRec & \brstar & \flare & \%SotA & \%ID \\
        \cmidrule(r){3-10} \cmidrule(l){11-19}
        
        \multirow{3}{1.1cm}{Pets}
        & Recall@10 & 0.077 & 0.093 & \textbf{0.116} & 0.104 & 0.101 & \underline{0.112} & \red{-3.45\%} & \green{+10.89\%}
            & 0.0499 & 0.0826 & 0.0647 & 0.0970 & 0.0937 & \underline{0.0997} & \textbf{0.1091} & \green{+12.47\%} & \green{+9.43\%} \\
        
        & nDCG@10   & 0.060 & 0.070 & \textbf{0.097} & 0.074 & 0.081 & \underline{0.092} & \red{-5.15\%} & \green{+13.58\%}
            & 0.0376 & 0.0590 & 0.0392 & 0.0574 & 0.0736 & \underline{0.0762} & \textbf{0.0817} & \green{+11.01\%} & \green{+7.22\%} \\
        
        & MRR       & 0.059 & 0.065 & \textbf{0.094} & 0.071 & 0.077 & \underline{0.088} & \red{-6.38\%} & \green{+14.29\%}
            & 0.0365 & 0.0549 & 0.0349 & 0.0503 & 0.0696 & \underline{0.0727} & \textbf{0.0775} & \green{+11.35\%} & \green{+6.60\%} \\
        
        \cmidrule(r){3-10} \cmidrule(l){11-19}

        \multirow{3}{1.1cm}{Office}
        & Recall@10 & 0.121 & 0.126 & 0.140 &  0.119 & \underline{0.141} & \textbf{0.148} & \green{+5.71\%} & \green{+4.96\%}
            & 0.0709 & 0.1063 & 0.0941 & \underline{0.1220} & 0.1213 & 0.1172 & \textbf{0.1240} & \green{+1.64\%} & \green{+5.80\%} \\
        
        & nDCG@10   & 0.097 & 0.092 & 0.114 & 0.091 & \underline{0.115} & \textbf{0.121} & \green{+6.14\%} & \green{+5.22\%}
            & 0.0545 & 0.0687 & 0.0607 & 0.0713 & \textbf{0.0976} & \underline{0.0935} & \textbf{0.0976} & \green{+0\%} & \green{+4.39\%} \\
        
        & MRR       & 0.093 & 0.085 & \underline{0.109} & 0.096 & \underline{0.109} & \textbf{0.115} & \green{+5.50\%} & \green{+5.50\%}
            & 0.0520 & 0.0603 & 0.0535 & 0.0597 & \textbf{0.0925} & \underline{0.0886} & \textbf{0.0925} & \green{+0\%} & \green{+4.40\%} \\

        \cmidrule(r){3-10} \cmidrule(l){11-19}
        
        \multirow{3}{1.1cm}{Arts}
        & Recall@10 & 0.124 & 0.133 & \textbf{0.161} &  0.140 & 0.144     & \underline{0.153} & \red{-4.97\%} & \green{+6.25\%}
            & 0.0799 & 0.1095 & 0.1052 & \textbf{0.1320} & 0.1140 & 0.1067 & \underline{0.1145} & \red{-13.26\%} & \green{+7.31\%} \\

        & nDCG@10   & 0.094 & 0.089 & \textbf{0.125} &  0.103 & 0.111     & \underline{0.121} & \red{-3.20\%} & \green{+9.01\%}
            & 0.0547 & 0.0662 & 0.0602 & 0.0729 & \textbf{0.0864} & 0.0787 & \underline{0.0822} & \red{-4.86\%} & \green{+4.45\%} \\

        & MRR       & 0.090 & 0.080 & \textbf{0.119} &  0.106 & 0.103     & \underline{0.114} & \red{-4.20\%} & \green{+10.68\%}
            & 0.0513 & 0.0582 & 0.0520 & 0.0615 & \textbf{0.0815} & 0.0739 & \underline{0.0766} & \red{-6.01\%} & \green{+3.65\%} \\

        \cmidrule(r){3-10} \cmidrule(l){11-19}
        
        \multirow{3}{1.1cm}{Games}
        & Recall@10 & 0.103 & 0.092 & 0.104 & 0.088 & \underline{0.110} & \textbf{0.117} & \green{+12.50\%} & \green{+6.36\%}
            & 0.0646 & 0.0724 & 0.0884 & \textbf{0.1074} & \underline{0.0986} & 0.0776 & 0.0980 & \red{-8.75\%} & \green{+26.29\%} \\

        & nDCG@10 & 0.063 & 0.058 & 0.068 & 0.053 & \underline{0.072} & \textbf{0.075} & \green{+10.29\%} & \green{+4.17\%}
            & 0.0334 & 0.0377 & 0.0419 & 0.0501 & \textbf{0.0595} & 0.0439 & \underline{0.0536} & \red{-9.92\%} & \green{+22.10\%}\\

        & MRR & 0.059 & 0.055 & 0.065 & 0.050 & \underline{0.067} & \textbf{0.069} & \green{+6.15\%} & \green{+2.99\%}
            & 0.0313 & 0.0349 & 0.0367 & 0.0427 & \textbf{0.0510} & 0.0394 & \underline{0.0476} & \red{-6.67\%} & \green{+20.81\%} \\
        
        \cmidrule(r){3-10} \cmidrule(l){11-19}

        \multirow{3}{1.1cm}{Music}
        & Recall@10 & 0.097 & \underline{0.112} & 0.105 & 0.111 & 0.103     & \textbf{0.112} & - & \green{+8.74\%}
            & 0.0915 & 0.0940 & 0.1048 & \textbf{0.1255} & 0.1158 & 0.1092 & \underline{0.1177} & \red{-6.22\%} & \green{+7.78\%}\\

        & nDCG@10   & 0.071 & 0.079 & \underline{0.083} & 0.079 & 0.078     & \textbf{0.085} & \green{+2.41\%} & \green{+8.97\%}
            & 0.0680 & 0.0596 & 0.0606 & 0.0709 & \textbf{0.0909} & 0.0848 & \underline{0.0881} & \red{-3.08\%} & \green{+3.89\%} \\

        & MRR       & 0.068 & 0.074 & \underline{0.081} & 0.076 & 0.073     & \textbf{0.081} & - & \green{+10.96\%}
            & 0.0658 & 0.0535 & 0.0526 & 0.0613 & \textbf{0.0864} & 0.0812 & \underline{0.0838} & \red{-3.01\%} & \green{+3.20\%} \\

        \cmidrule(r){3-10} \cmidrule(l){11-19}

        \multirow{3}{1.1cm}{Scientific}
        & Recall@10 & 0.106 & \underline{0.126} & \textbf{0.145}  & 0.080 & 0.113     & 0.119 & \red{-17.93\%} & \green{+5.31\%}
            & 0.0636 & 0.1058 & 0.0727 & \textbf{0.1181} & \underline{0.1124} & 0.0829 & 0.0925 & \red{-21.68\%} & \green{+11.58\%} \\

        & nDCG@10   & 0.079 & \underline{0.086} & \textbf{0.103} & 0.045 & 0.080     & 0.084 & \red{-18.45\%} & \green{+5.00\%}
            & 0.0411 & 0.0639 & 0.0414 & \underline{0.0644} & \textbf{0.0788} & 0.0550 & 0.0603 & \red{-23.48\%} & \green{+9.64\%} \\

        & MRR       & 0.076 & \underline{0.079} & \textbf{0.095} & 0.039 & 0.074     & 0.076 & \red{-20.00\%} & \green{+2.70\%}
            & 0.0389 & \underline{0.0570} & 0.0380 & 0.0555 & \textbf{0.0730} & 0.0503 & 0.0547 & \red{-25.07\%} & \green{+8.75\%} \\

        \bottomrule
    \end{tabular}}
    \label{tab:baselines_res}
\end{table*}

\subsection{RQ1: \flare Performance on Existing Benchmarks}
\label{sec:small_datasets}

To assess the efficacy of \flare, we first want to ensure that the baseline model is fully-tuned. We evaluate our \brstar (tuned Bert4Rec) and \flare models on commonly used datasets in the literature. Reference performance values of prior models are quoted from Recformer~\cite{recformer} and CALRec~\cite{Li2024-cx} which use the same evaluation methodology and data preprocessing.

Table~\ref{tab:baselines_res} shows the performance results, using metrics common to those reported in Recformer \& CALRec (e.g. mean reciprocal rank, nCDG, and recall \cite{Zangerle2023-vo}). \textbf{Bold} values indicate the best result, and \underline{underline} values indicate the second best. The ``\%SotA'' column denotes the performance difference between \flare and the best prior model reported in the literature. The ``\%ID'' column is the difference between \flare and \brstar.

The \brstar baseline, which does not leverage text, (a) significantly outperforms the original Bert4Rec baselines; (b) beats prior work of Recformer on 2 out of 6 datasets (``Office'' and ``Games''); and (c) is generally higher than S$^3$Rec in all datasets except the Music dataset.

\flare outperforms (a) S$^3$Rec on all datasets; (b) all prior models on Recformer-processed Office, Games, and Music datasets; and (c) all prior models on CALRec-processed Pets and Office datasets. It is second-best on most other datasets and only modestly under-performs on CALRec-processed Music. One possible reason for \flare's under-per\-for\-mance on the remaining datasets is that Recformer and CALRec use two-phase fine-tuning, where the underlying language model is tuned on a large corpus of examples from the dataset. In contrast, \flare uses a frozen mT5 model, with no pre-training on the Amazon datasets. We believe tuning in-domain LMs or using in-domain content embeddings may improve performance.

Comparing between our own models, \flare outperforms \brstar on all datasets, suggesting that the combination of IDs and text does improve performance. Notably, the gains enabled by modeling text and IDs extend to the smallest Amazon Product Reviews datasets, such as the Scientific and Music datasets, containing just 5k and 10k items, respectively.

In summary, our experiments confirm (1) the performance gains of combining IDs and text, even with an updated, stronger \brstar baseline; and (2) that \flare performs reasonably well compared to previously reported baselines.

\subsection{RQ2: Performance on Large Datasets}

\begin{table}
    \setlength{\tabcolsep}{2.2pt}
    \centering
    \caption{Performance of \flare and \brstar (tuned Bert4Rec) scaling on Clothing, a dataset with over 370k items. \flare achieves improved performance when increasing the model size. The \% columns show the performance improvement of the best model over Recformer.}

    \resizebox{0.49\textwidth}{!}{
    \begin{tabular}{l c c c c c c c c c}
        \toprule
        &  & \multicolumn{4}{c}{\brstar} & \multicolumn{4}{c}{Flare} \\

        \cmidrule(r){3-6} \cmidrule{7-10}

        Metrics & Recformer & Small & Base & Large & \% & Small & Base & Large & \% \\
        \cmidrule(r){1-1} \cmidrule(lr){2-2} \cmidrule(r){3-6} \cmidrule{7-10}

        Recall@1  & - & 0.079 & 0.127 & 0.128 & - & 0.123 & \underline{0.129} & \textbf{0.130} & - \\
        Recall@5  & - & 0.106 & \underline{0.138} & \underline{0.138} & - & 0.137 & \textbf{0.142} & \textbf{0.142} & - \\
        Recall@10 & 0.135 & 0.112 & 0.143 & 0.143 & +5.93\% & 0.143 & \textbf{0.148} & \underline{0.147} & +9.63\% \\
        nDCG@10   & 0.123 & 0.096 & 0.134 & \underline{0.135} & +9.76\% & 0.132 & \textbf{0.138} & \textbf{0.138} & +12.20\% \\
        MRR       & 0.120 & 0.091 & \underline{0.133} & \underline{0.133} & +10.83\% & 0.130 & \textbf{0.136} & \textbf{0.136} & +13.33\% \\
        \bottomrule
    \end{tabular}}
    \label{tab:large_vocab}
\end{table}

Many production applications of recommender systems are faced with vocabulary sizes which are orders of magnitude larger than the datasets above. As such, a question arises whether performance gains scale with larger vocabulary datasets.

To study this problem, we use the Clothing, Shoes and Jewelry category from Amazon Product Reviews (376K Items) and report the performance of \flare models using different model sizes (Table~\ref{tab:large_vocab}). Given the larger candidate set, we also include Recall@1 and Recall@5. We selected three representative model sizes reported as (i) ``Small'' with 2 layers, 2 heads, 64 model dimensions, and 256 hidden dimensions; (ii) ``Base'' with 8 layers, 16 heads, 768 model dimensions, and 3072 hidden dimensions; and (iii) ``Large'' with 32 layers, 32 heads, 768 model dimensions, and 3072 hidden dimensions. For all \flare models, the perceiver model size is 8 layers, 16 heads, and 8 latents. The results show that increasing the model size improves performance, up to a point of diminishing returns.

We benchmark \flare against two baselines: \brstar and Recformer. \brstar here is a Bert4Rec model tuned for the Clothing dataset. Comparing the \flare models with the \brstar models confirms that the addition of text helps, independently of the model size, especially on the smaller, more constrained models.

The Recformer paper did not evaluate performance on the Clothing dataset. Thus we downloaded Recformer's published pretrained checkpoints and ran the fine-tuning pipeline code. Before fine-tuning Recformer on our Clothing dataset, we confirmed our ability to successfully reproduce Recformer's published results on the Scientific dataset, within -0.5\% and +3\% across runs. When fine-tuning on Clothing, one notable set of changes required is the fine-tuning data sampling rate and batch size. For the small datasets, Recformer randomly samples 20\% of the dataset with a batch size of 16. For Clothing, we could only sample 18.18\% with a batch size of 8 given memory constraints of the H100 80GB GPU used for training.

For their small dataset evaluation, Recformer pretrains on a set of datasets that is disjoint from the set used for fine-tuning. Recformer's pretrained checkpoint includes Clothing, so fine-tuning that checkpoint on Clothing would depart from Recformer's evaluation methodology. Though, we expect that because the model sees the Clothing dataset multiple times, the performance in this setting is ``optimistic'' and thus serves as a more challenging baseline. Even so, the most performant \flare model outperforms on all metrics and establishes a new baseline for this dataset (Table~\ref{tab:large_vocab}).

\subsection{RQ3: Critiquing}
\label{sec:critiquing}
We evaluate the ability for \flare to utilize textual signals through the critiquing task described below. We use the ``Office'' and ``Clothing'' datasets from Amazon Product Reviews, following the preprocessing steps as described for the ``Clothing'' dataset (Section~\ref{sec:clothing_dataset}).
We formulate Amazon Product Reviews for this task by using each item's ``category'' attribute as the critique. Each provided category string consists of a hierarchical order of categories where each category term is progressively more precise, delimited by ``-''. This characteristic enables us to evaluate two different levels of critiquing:
\begin{itemize}
    \item \textbf{Precise Critiquing:} Critiquing with a specific category using the first four item categories (Ex: \textit{Office Products - Office and School Supplies - Printer Ink \& Toner - Inkjet Printer Ink})
    \item \textbf{Broad Critiquing:} Critiquing with a broad category using the first two item categories (Ex: \textit{Office Products - Office and School Supplies})
\end{itemize}
By including these text based hints, we expect that a model that effectively leverages text semantics will achieve greater performance under Precise Critiquing, slightly less performance under Broad Critiquing, and worst performance without critiquing.

\subsubsection{Results}
\begin{table}
    \centering
    \caption{\flare's performance when utilizing precise (4 categories) and broad (2 categories) textual critiquing. \flare achieves stronger performance under both forms of critiquing, demonstrating that it can leverage textual critiques, with some room for improvement with broad critiquing.}    
    \begin{tabular}{l l c c c}
        \toprule
        Dataset  & Metrics & Precise & Broad & None  \\
        \midrule
        \multirow{5}{1.7cm}{Office}
        & Recall@1 & 0.213 & 0.138 & 0.124   \\
                                & Recall@5 & 0.355 & 0.180 & 0.158 \\
                                & Recall@10 & 0.437 & 0.203 & 0.173 \\
                                & nDCG@10 & 0.313 & 0.168 & 0.147 \\
                                & MRR & 0.286 & 0.161 & 0.141 \\
        \midrule
        \multirow{5}{1.8cm}{Clothing}
        & Recall@1 & 0.167 & 0.135 & 0.129  \\
                                  & Recall@5 & 0.229 & 0.151 & 0.142 \\
                                  & Recall@10 & 0.272 & 0.161 & 0.148 \\
                                  & nDCG@10 & 0.212 & 0.147 & 0.138 \\
                                  & MRR & 0.202 & 0.144 & 0.136 \\
        \bottomrule
    \end{tabular}
    \label{tab:critiquing}
\end{table}
Results on Precise and Broad Critiquing on the Office and Clothing datasets are shown in Table~\ref{tab:critiquing}. Indeed, Precise Critiquing performance is higher than that of Broad Critiquing, and both reach higher performance than not using critique text. Compared to the baseline without critiquing, Precise Critiquing obtained +153\% and +84\% higher Recall@10 on the Office and Clothing datasets. The vastly improved performance achieved with critiquing suggests that \flare can leverage textual instruction.

Notably, we observe that Broad Critiquing only improves performance by +17\% and +9\% Recall@10 on the same datasets, a much smaller increase compared to Precise Critiquing. We hypothesize that the reduced performance gain is due to the lack of specificity of the critiquing string, offering a challenge for future work. Appendix~\ref{sec:appendix_example_critiquing} details further qualitative analysis of critiquing outputs.

\subsection{RQ4: Out-of-Sequence Critiquing}
We now explore whether critiquing strings can steer the model towards out-of-domain categories from the original user sequence.
For this, we construct a variant of the Office test set where we randomly mutate the last item critiquing category to another valid critiquing category within the dataset. We then evaluate our best pre-trained \flare model trained under precise critiquing on that modified dataset.
Using category mutations, we evaluate \flare's ability to follow the critique, even if it is increasingly out-of-context given the user's history. For example, rather than the ground truth of shopping for a desk chair, supported by a user sequence, we mutate the last critique and query the model to recommend a large industrial-grade TV. In this case, an effective model would handle the critique request despite being increasingly out-of-domain of the original, unmodified user sequence.

\subsubsection{Category nDCG Metric}
Since there is no longer ground truth items for this task, we propose Cat-nDCG, a variant of nDCG, to quantify whether the model's predictions are aligned with the mutated critiquing string.

We define Cat-nDCG as follows. Let $S'$ be the ``target'' critiquing string given as input to the model. Let $K$ be the set of retrieved items sorted by confidence, and $S_k = \{s_1, s_2, s_3, s_4\}$ be the set of hierarchically ordered categories associated with item $k \in K$. We then define a retrieved item $k$'s relevance $rel_k$ as
\begin{equation}
rel_k = |S_k \cap S'|
\end{equation}
the number of categories in the predicted item $k$ overlapping with the critiquing string $S'$. As the categories in $S$ and $S'$ are hierarchical, an item that matches $s_j$ in $S'$ will also have matched all $s_i, i < j$. For instance, a retrieved item matching all four categories will have a relevance of 4, whereas an item only matching two parent categories ($s_1, s_2$ but not $s_3, s_4$) will have a relevance of 2. We then apply nDCG \cite{Jarvelin2002-pg, jarvelin2000ir} as usual using these relevance values. As a result, Cat-nDCG evaluates rank-wise whether the model retrieves items matching the given critique category.

\subsubsection{Results}
\begin{table}
    \centering
    \caption{Performance of \flare's category alignment when mutating the critiquing category string to another valid category on Office. \flare's predictions are nearly entirely within the desired category, including when the target category vastly differs from the user's prior actions.}    
    \begin{tabular}{l c}
    \toprule
        Mutated Categories & Cat-nDCG@10 \\
    \midrule
        Level 4 ($s_4$) & 0.944\\
        Level 3 ($s_3, s_4$)& 0.917\\
        Level 2 ($s_2, s_3, s_4$) & 0.890\\
    \bottomrule
    \end{tabular}
    \label{tab:changing_critique}
\end{table}

To construct the input critique $S'$, we take an original category string $C_k = \{c_1, c_2, c_3, c_4\}$ for an item $k$. We consider candidate set $C^{j}_{k}$ obtained by randomly mutating each of the component categories $c_l \in C_k, j \leq l \leq 4$ such that the resulting string maps to a valid category in the dataset. Then we randomly sample a critiquing string from this set, which serves as $S'$. To ensure that mutated categories are high quality, we require that the chosen category $S'$ has at least five items in the dataset. As we decrease $j$, the mutated categories move further from the original string $C_k$, therefore becoming increasingly more challenging to adapt.

Our results setting $j = 4, 3, 2$, referred to as Level $j$, is shown in Table~\ref{tab:changing_critique}. Starting with Level 4 mutations, \flare achieves a Cat-nDCG@10 of 0.944, indicating that most retrieved items follow the critiquing string and are highly ranked. This finding aligns with our qualitative analysis in Appendix~\ref{sec:appendix_example_critiquing},
suggesting that \flare can generalize to outside-history categories. Notably, when increasing to Level 2 mutations, \flare's Cat-nDCG@10 score drops to 0.890. This may be due to a conflict between the user's historical signals and the desired category value, suggesting an area for future work.

\subsection{Ablation Study}

We perform ablation studies to demonstrate the impact of various components in our proposed model. The results are shown in Table~\ref{tab:ablation} for the Clothing dataset with the Base model size of 8 layers, 16 heads, 768 model dimensions, and 3072 hidden dimensions.

To evaluate the impact of the Perceiver model, we experiment with using the EOS token embedding directly from a text sentence. While we find that the impact of Perceiver is less than expected, it enables future applications involving multi-modal inputs.

Our use of bidirectional masking and contrastive loss contributes to \flare's performance. Removing these features resulted in a $\sim-8\%$ and $\sim-6\%$ drop in Recall@5, respectively. In particular, the most significant decrease in MRR came from removing bidirectional masking, despite evaluation always masking the last item. We suspect that the drop in ranking performance is due to bidirectional masking reinforcing the model to leverage signals from across the sequence. Further, bidirectional masking is a more strenuous training task as multiple items can be masked in a sequence rather than just the last item.

Finally, we also evaluate the impact of removing duplicates in the user's sequence. The model's performance does not decrease significantly, with Recall@5 only dropping by $\sim 1\%$. This suggests that the model's performance is unlikely due to copying previous items in the sequence, at least for the Clothing dataset.

\begin{table}
    \centering
    \caption{\flare ablation study using the Clothing dataset.}
    \resizebox{0.48\textwidth}{!}{\begin{tabular}{l c c c c}
        \toprule
        Method  & Recall@5 & Recall@10 & nDCG@10 & MRR \\
        \midrule
        \flare & 0.142 & 0.148 & 0.138 & 0.136 \\
        w/o Text  & 0.138 & 0.143 & 0.134 & 0.133 \\
        w/o Perceiver & 0.140 & 0.145 & 0.135 & 0.133 \\
        w/o Bidirectional Masking & 0.131 & 0.134 & 0.128 & 0.126 \\
        w/o Contrastive Loss & 0.134 & 0.145 & 0.136 & 0.134 \\
        w/o Duplicates & 0.140 & 0.146 & 0.136 & 0.134 \\
        \bottomrule
    \end{tabular}}
    \label{tab:ablation}
\end{table}

\section{Conclusion and Future Work}
\label{sec:conclusion}
We introduce \flare, a hybrid model recommender to incorporate item IDs with contextual text information. In \brstar, we tune Bert4Rec and demonstrate that the tuned model outperforms previous Bert4Rec baselines, as well as state-of-the-art models on two datasets. \flare shows additional performance gains over the new \brstar baseline, suggesting a benefit to hybrid recommenders. We also extend existing benchmarks by introducing a baseline for larger item count domains, where \flare outperforms the \brstar and Recformer baselines. We believe these results may serve as new baselines for future work.

Our results suggest hybrid methods that incorporate both direct item representation, along with item context like text, is a promising approach for quality. Regarding text, one area for future work is investigating the impact on recommendation performance from different types of language models, and their fine-tuning strategies, when used to encode text metadata. Another consideration is the substantial high memory cost of ID embedding tables at scale. Extending models to handle very large vocabularies---beyond the datasets used in this paper---may be considerably challenging. We plan to focus on the impact of different vocabulary sizes and investigate new techniques to handle larger item counts.

Finally, we propose the use of critiquing as an evaluation methodology to measure how well a model interprets and acts on textual critique strings. Though using the category feature in the Amazon Product Reviews dataset as a proxy for a user-supplied critique is arguably contrived, we view it as a modest step towards natural-language critiquing that might occur in a conversational recommender. We plan to focus on expanding the methodology to use more ``realistic'' critiques in future work.

\begin{acks}
We thank James Ren, Ajit Apte, Dima Kuzmin, and Santiago Ontanon for their expert advice and feedback.
\end{acks}

\bibliographystyle{ACM-Reference-Format}

\bibliography{main}

\appendix
\balance
\section{Performance on Unseen Users}

\begin{table}[b]
    \centering
    \caption{Performance when evaluating on unseen users versus on the unseen last item of each user.}
    \resizebox{0.49\textwidth}{!}{\begin{tabular}{l l  c c c c c c c c}
                \toprule
                &         & \multicolumn{2}{c}{Unseen Users} & \multicolumn{2}{c}{Unseen Last Item} \\
                \cmidrule(lr){3-4} \cmidrule(l){5-6}
        Dataset & Metrics & \brstar & \flare & \brstar & \flare  \\
        \midrule
                \multirow{5}{2cm}{Office}
                                & Recall@1 & 0.101 & 0.121  & 0.099 & 0.124 \\
                                & Recall@5 & 0.142 & 0.155 & 0.145 & 0.158 \\
                                & Recall@10 & 0.155 & 0.169 & 0.160 & 0.173 \\
                                & nDCG@10 & 0.128 & 0.145 & 0.128 & 0.147  \\
                                & MRR & 0.121 & 0.140 & 0.121 & 0.142  \\
        \midrule
                \multirow{5}{2cm}{Clothing}
                                & Recall@1 & 0.126 & 0.129 & 0.127 & 0.129 \\
                                & Recall@5 & 0.136 & 0.141 & 0.138 & 0.142 \\
                                & Recall@10 & 0.140 & 0.146 & 0.143 & 0.148 \\
                                & nDCG@10 & 0.133 & 0.137 & 0.135 & 0.138  \\
                                & MRR & 0.131 & 0.135 & 0.133 & 0.136  \\
        \bottomrule
    \end{tabular}}
    \label{tab:unseen_users}
\end{table}

\begin{table*}
    \centering
    \caption{Hyperparameters for \brstar.}

    \begin{tabular}{l c c c c c c  c c c c c c}
    \toprule
        & \multicolumn{6}{c}{Recformer-Processed Datasets} & \multicolumn{6}{c}{CALRec-Processed Datasets} \\
        \cmidrule(r){2-7} \cmidrule(l){8-13}
    Params          & Games & Office    & Scientific    & Music & Arts  & Pets  & Games & Office    & Scientific    & Music & Arts  & Pets  \\
    \cmidrule(r){1-1} \cmidrule(r){2-7} \cmidrule(l){8-13}
    \# Layers       & 2     & 2         & 4             & 8     & 2     & 2     & 4     & 4         & 2             & 4     & 2     & 4 \\
    \# Heads        & 2     & 2         & 16            & 8     & 8     & 16    & 16    & 16        & 8             & 16    & 8     & 16 \\
    Model Dims      & 64    & 64        & 512           & 256   & 768   & 1024  & 1024  & 1024      & 768           & 1024  & 768   & 1024 \\
    Hidden Dims     & 256   & 256       & 2048          & 1024  & 3072  & 4096  & 4096  & 4096      & 3072          & 4096  & 3072  & 3072 \\
    Learning Rate   & 1e-3  & 1e-3      & 1e-4          & 1e-5  & 1e-4  & 1e-4  & 1e-4  & 1e-4      & 1e-4          & 1e-4  & 1e-4  & 1e-4  \\
    Batch           & 1     & 1         & 16            & 16    & 16    & 32    & 32    & 32        & 32            & 32    & 32    & 32 \\
    Steps           & 50K   & 50K       & 50K           & 50K   & 50K   & 10K   & 1.6K  & 1.9K      & 800           & 900   & 1.5K  & 1.9K \\
    \bottomrule
    \end{tabular}
    \label{tab:hparams-bert4rec}
\end{table*}

\begin{table*}
    \centering
    \caption{Hyperparameters for for \flare.}
    \begin{tabular}{l c c c c c c  c c c c c c}
    \toprule
        & \multicolumn{6}{c}{Recformer-Processed Datasets} & \multicolumn{6}{c}{CALRec-Processed Datasets} \\
        \cmidrule(r){2-7} \cmidrule(l){8-13}
        Params              & Games & Office    & Scientific    & Music & Arts  & Pets  & Games & Office    & Scientific    & Music & Arts  & Pets\\
    \cmidrule(r){1-1} \cmidrule(r){2-7} \cmidrule(l){8-13}
        \# Layers           & 8     & 2         & 2             & 2     & 4     & 2     & 8     & 4         & 2             & 2     & 4     & 4 \\
        \# Heads            & 16    & 4         & 8             & 2     & 4     & 8     & 4     & 4         & 2             & 2     & 8     & 4 \\
        Model Dims          & 1024  & 768       & 256           & 1024  & 512   & 256   & 256   & 128       & 1280          & 128   & 128   & 512 \\ 
        Hidden Dims         & 4096  & 3072      & 1024          & 4096  & 2048  & 1024  & 1024  & 512       & 5120          & 512   & 512   & 2048 \\
        Learning Rate       & 1e-4  & 1e-4      & 1e-4          & 1e-5  & 1e-4  & 1e-4  & 1e-4  & 1e-4      & 1e-4          & 1e-4  & 1e-4  & 1e-4 \\
        Batch               & 2     & 16        & 8             & 2     & 8     & 16    & 32    & 32        & 64            & 32    & 32    & 16 \\
        Steps               & 5K    & 5K        & 5             & 25K   & 10K   & 25K   & 9.5K  & 7K        & 5.5K          & 7K    & 10K   & 8K \\
        Perceiver Heads     & 16    & 16        & 8             & 8     & 2     & 8     & 8     & 8         & 8             & 8     & 8     & 8 \\
        Perceiver Layers    & 8     & 2         & 2             & 8     & 2     & 2     & 4     & 4         & 4             & 4     & 4     & 4 \\
        Perceiver Latents   & 2     & 8         & 2             & 4     & 2     & 2     & 4     & 4         & 4             & 4     & 4     & 4 \\
        Weight Decay        & 1e-3  & 1e-2      & 1e-3          & 1e-3  & 1e-3  & 1e-3  & 1e-3  & 1e-3      & 1e-3          & 1e-3  & 1e-3  & 1e-3 \\
    \bottomrule
    \end{tabular}
    \label{tab:hparams-flare}
\end{table*}

In this paper, we employ the common evaluation paradigm of using the last two items for each user as validation and hold-out sets. Thus, the model processes similar user sequences during training and evaluation. To study how well the model generalizes to unseen users in training, we also evaluate the model using training and test splits with non-overlapping user sequences. We construct an 80\%/10\%/10\% split of users to comprise the train, test and validation sets respectively. To match the sequence length of the unseen-last-item setting, we withhold the last two items from each training example. We use the same preprocessing approach as the Clothing dataset (Section~\ref{sec:clothing_dataset}). Table~\ref{tab:unseen_users} shows that \brstar and \flare  achieve similar results in both evaluation settings, suggesting that the model generalizes well to unseen users.

\section{Model Hyperparameters}
\label{sec:appendix_hyperparameters}
Tables~\ref{tab:hparams-bert4rec} and \ref{tab:hparams-flare} summarize the current-best discovered \brstar and \flare model hyperparameters.

\section{Example Traces with Critiquing}
\label{sec:appendix_example_critiquing}
This section highlights a few recommendation traces using critiquing. The top segment of Table~\ref{tab:traces} shows an example input item sequence, taken from the Amazon Reviews Offices dataset. The sequence contains four items of various office supplies comprising a letter organizer (repeated), printer ink, and staples.

For comparison, the subsequent segments in Table~\ref{tab:traces} show recommendations for that input sequence using: (1) \brstar, where each item is represented as an item ID only; (2) \flare, where each item is an ID plus descriptive text (e.g. title); and (3) \flare, with a critique input at the end of the sequence. The critique shown is based on the category of the held-out labeled item (\texttt{\small T50 Heavy Duty Staples [Set of 2]}). As expected, the variety and relevance of item categories narrows when comparing \brstar to \flare to \flare-with-critiquing. For instance, the \flare recommendations contain items in the \texttt{\small Inkjet Printer Ink} and \texttt{\small Staples}, mimicking the items in the input sequence, while \flare with critiquing returns homogeneous results within the requested category.

The critiquing example above uses the category of the held-out label for the input. Table~\ref{tab:alt_traces} demonstrates recommendations for the same input sequence when the critique is out of that input's sequence distribution. The table segments are recommendations from critiques using: (1) a category already represented in the input sequence; (2) a category not represented in the input sequence; (3) a broad category; and (4) a narrow category.\\

\begin{table*}[]
    \centering
    \small
    \begin{tabular}{p{0.5cm}|p{0.48\textwidth}|p{0.42\textwidth}}
    \toprule
    \multicolumn{3}{l}{\textbf{User Input History}} \\
    \hline
    Seq. & Title & Category \\
    \hline
    1 & Spectrum Diversified Scroll Mail and Letter Organizer, Wall Mount, Black &
    Office Products - Office\& School Supplies - Forms, Recordkeeping \& Money Handling - Key Cabinets, Racks \& Holders \\

    2 & Spectrum Diversified Scroll Mail and Letter Organizer, Wall Mount, Black &
    Office Products - Office \& School Supplies - Forms, Recordkeeping \& Money Handling - Key Cabinets, Racks \& Holders  \\

    3 & Canon PGI-250/CLI-251 Combo Pack &
    Office Products - Office \& School Supplies - Printer Ink \& Toner - Inkjet Printer Ink \\

    4 & Arrow Fastener 506 Genuine T50 3/8-Inch Staples, 1250-Pack &
    Office Products - Office \& School Supplies - Staplers \& Punches - Staples \\

    \midrule
    \midrule
    \multicolumn{3}{l}{\textbf{Recommendations using \brstar}} \\
    \hline
    Rank & Title & Category \\
    \hline

    1 & Scotch Thermal Laminator 2 Roller System (TL901C) &
    Office Products - Office Electronics - Presentation Products - Laminators \\

    2 & Spectrum Diversified Scroll Mail and Letter Organizer, Wall Mount, Black &
    Office Products - Office \& School Supplies - Forms, Recordkeeping \& Money Handling - Key Cabinets, Racks \& Holders \\

    3 & Scotch Thermal Laminating Pouches, 8.9 x 11.4-Inches, 3 mil thick, 20-Pack (TP3854-20), Clear &
    Office Products - Office Electronics - Presentation Products - Laminating Supplies \\

    4 & Quartet Dry Erase Board, Whiteboard / White Board, Magnetic, 17'' x 23'', Assorted Frame Colors - Color Will Vary, 1 Board (MWDW1723M) &
    Office Products - Office \& School Supplies - Presentation Boards - Dry Erase Boards \\

    5 & Sakura Pigma 30062 Micron Blister Card Ink Pen Set, Black, Ass't Point Sizes 6CT Set &
    Office Products - Office \& School Supplies - Writing \& Correction Supplies - Pens \& Refills - Calligraphy Pens \\

    \midrule
    \midrule
    \multicolumn{3}{l}{\textbf{Recommendations using \flare (item ID + textual context)}} \\
    \hline
    Rank & Title & Category \\
    \hline

    1 & Sharpie 37001 Permanent Markers, Ultra Fine Point, Black, 12 Count &
    Office Products - Office \& School Supplies - Writing \& Correction Supplies - Markers \& Highlighters - Permanent Markers \\

    2 & Canon PGI-250XL Pigment Black Ink, Twin Value Pack, Compatible to MG5520, MG6620, MG5420, ... &
    Office Products - Office \& School Supplies - Printer Ink \& Toner - Inkjet Printer Ink \\

    3 & Arrow Fastener 506 Genuine T50 3/8-Inch Staples, 1250-Pack &
    Office Products - Office \& School Supplies - Staplers \& Punches - Staples \\

    4. & AT\&T CL2940 Corded Phone with Caller ID/Call waiting, Speakerphone, XL Tilt Display, ... &
    Office Products - Office Electronics - Telephones \& Accessories - Landline Phones - Corded Telephones \\

    5 & Pilot G2 Retractable Premium Gel Ink Roller Ball Pens Fine Pt (.7) Dozen Box Red ; Retractable, Refillable \& Premium ... &
    Office Products - Office \& School Supplies - Writing \& Correction Supplies - Pens \& Refills - Rollerball Pens - Gel Ink Rollerball Pens \\

    \midrule
    \midrule
    \multicolumn{3}{l}{\textbf{Recommendations using \flare, with label's category as critique: \emph{Office Products - Office \& School Supplies - Staplers \& Punches}}} \\
    \hline
    Rank & Title & Category \\
    \hline

    1 & T50 Heavy Duty Staples [Set of 2] &
    Office Products - Office \& School Supplies - Staplers \& Punches \\

    2 & Onotio Heavy Duty 100 Sheet High Capacity Office Desk Stapler with 1000 Box Staples &
    Office Products - Office \& School Supplies - Staplers \& Punches \\

    3 & Swingline Stapler, Optima 70, Desktop Stapler, 70 Sheet Capacity, Reduced Effort, Half Strip, Silver (87875) &
    Office Products - Office \& School Supplies - Staplers \& Punches \\

    4 & 1 X Flat Clinch Staples Mini Box of 1000 by MAX No.10 &
    Office Products - Office \& School Supplies - Staplers \& Punches \\

    5 & Swingline Automatic Stapler, Breeze, 20 Sheet Capacity, Battery Powered, Color Selected For You (S7042131) &
    Office Products - Office \& School Supplies - Staplers \& Punches \\
    \bottomrule
    \end{tabular}
    \caption{Example input sequence and recommendations from Bert4Rec, \flare, and \flare with a category-based critiquing input. The critique is the category of the held-out labeled item.}
    \label{tab:traces}
\end{table*}

\begin{table*}[]
    \small
    \centering

    \begin{tabular}{p{0.5cm}|p{0.48\textwidth}|p{0.42\textwidth}}
    \toprule

    \multicolumn{3}{l}{\textbf{User Input History}} \\
    \hline
    Seq. & Title & Category \\
    \hline
    1 & Spectrum Diversified Scroll Mail and Letter Organizer, Wall Mount, Black &
    Office Products - Office\& School Supplies - Forms, Recordkeeping \& Money Handling - Key Cabinets, Racks \& Holders \\

    2 & Spectrum Diversified Scroll Mail and Letter Organizer, Wall Mount, Black &
    Office Products - Office \& School Supplies - Forms, Recordkeeping \& Money Handling - Key Cabinets, Racks \& Holders  \\

    3 & Canon PGI-250/CLI-251 Combo Pack &
    Office Products - Office \& School Supplies - Printer Ink \& Toner - Inkjet Printer Ink \\

    4 & Arrow Fastener 506 Genuine T50 3/8-Inch Staples, 1250-Pack &
    Office Products - Office \& School Supplies - Staplers \& Punches - Staples \\

    \midrule
    \midrule
    \multicolumn{3}{l}{\textbf{Recommendations with alternate in-history category as critique: \emph{Forms, Recordkeeping \& Money Handling - Key Cabinets, Racks \& Holders}}} \\
    \hline
    Rank & Title & Category \\
    \hline
    1 & Spectrum Diversified Scroll Mail and Letter Organizer, Wall Mount, Black &
    Office Products - Office \& School Supplies - Forms, Recordkeeping \& Money Handling - Key Cabinets, Racks \& Holders \\

    2 & Lucky Line Designer Pattern Key Identifier Caps, 4 Pack (16304) &
    Office Products - Office \& School Supplies - Forms, Recordkeeping \& Money Handling - Key Cabinets, Racks \& Holders \\

    3 & KeyGuard SL-8548-U Dual Access Combination Key Cabinet With Chrome 4-Dial Combi-Cam Ultra - 48 Hook &
    Office Products - Office \& School Supplies - Forms, Recordkeeping \& Money Handling - Key Cabinets, Racks \& Holders - Cabinets \\

    4 & Master Lock Small Lock Box with 20 Key Capacity, 7131D, Beige &
    Office Products - Office \& School Supplies - Forms, Recordkeeping \& Money Handling - Key Cabinets, Racks \& Holders - Cabinets \\

    5 & Security Large Magnetic Hide-A-Key Holder for Over-Sized Keys - Extra-Strong Magnet &
    Office Products - Office \& School Supplies - Forms, Recordkeeping \& Money Handling - Key Cabinets, Racks \& Holders - Cabinets \\

    \midrule
    \midrule
    \multicolumn{3}{l}{\textbf{Recommendations with outside-history category as critique:}} \\
    \multicolumn{3}{l}{\textbf{\emph{Office Products - Office Electronics - PDAs, Handhelds \& Accessories - PDA \& Handheld Accessories}}} \\
    \hline
    Rank & Title & Category \\
    \hline

    1 & Kyocera Kona Black (Virgin Mobile) &
    Office Products - Office Electronics - PDAs, Handhelds \& Accessories - PDA \& Handheld Accessories \\

    2 & iPad Pro 9.7 Case, Apple iPad Pro 9.7 Folio Case, roocase Dual View Pro PU Leather Folio Stand Case for Apple iPad PRO 9.7-inch (2016), Black - NOT Compatible with iPad PRO 10.5 and 12.9, Black &
    Office Products - Office Electronics - PDAs, Handhelds \& Accessories - PDA \& Handheld Accessories \\

    3 & Saunders 00497 RhinoSkin Aluminum Hardcase for Palm Tungsten TX/T5 &
    Office Products - Office Electronics - PDAs, Handhelds \& Accessories - PDA \& Handheld Accessories \\

    4 & ZAGG Folio Case with Backlit Bluetooth Keyboard for Apple iPad Mini Purple-Black (1st Generation Only) Backlit Keyboard and Case &
    Office Products - Office Electronics - PDAs, Handhelds \& Accessories - PDA \& Handheld Accessories \\

    5 & HQRP Replacement Battery Palm Tungsten E2 PDA + Screwdriver + HQRP Universal Screen Protector &
    Office Products - Office Electronics - PDAs, Handhelds \& Accessories - PDA \& Handheld Accessories \\

    \midrule
    \midrule
    \multicolumn{3}{l}{\textbf{Recommendations with the label's broadened category as critique: \emph{Office Products - Office \& School Supplies}}} \\
    \hline
    Rank & Title & Category \\
    \hline
    1 & Canon CLI-8M Magenta Ink Tank &
    Office Products - Office \& School Supplies \\

    2 & Swingline Stapler, Optima 25, Full Size Desktop Stapler, 25 Sheet Capacity, Reduced Effort, Blue/Gray (66404) &
    Office Products - Office \& School Supplies \\

    3 & Mead \#10 Envelopes, Press-It Seal-It, White, 50/Box  (75024) &
    Office Products - Office \& School Supplies \\

    4 & Columbian \#10 Envelopes, Security Tinted,4-1/8'' x 9-1/2'', White, 500 Per Box, 2-PACK (CO128) &
    Office Products - Office \& School Supplies \\

    5 & Five Star Filler Paper, Graph Ruled Paper, 100 Sheets/Pack, 11'' x 8-1/2'', Reinforced, Loose Leaf, White (17016) &
    Office Products - Office \& School Supplies \\

    \midrule
    \midrule
    \multicolumn{3}{l}{\textbf{Recommendations with a the label's critique, made more specific: \emph{Office Products - Office \& School Supplies - Staplers \& Punches - Staples}}} \\
    \hline
    Rank & Title & Category \\
    \hline

    1 & Arrow Fastener 506 Genuine T50 3/8-Inch Staples, 1250-Pack &
    Office Products - Office \& School Supplies - Staplers \& Punches - Staples \\

    2 & Swingline Staples, S.F. 4, Premium, 1/4'' Length, 210/Strip, 5000/Box, 1 Box (35450) &
    Office Products - Office \& School Supplies - Staplers \& Punches - Staples \\

    3 & Swingline Staples, Standard, 1/4'' Length, 210/Strip, 5000/Box, 5 Pack (35101) &
    Office Products - Office \& School Supplies - Staplers \& Punches - Staples \\

    4 & Swingline Staples, Standard, 1/4'' Length, 210/Strip, 5000/Box, 1 Box (35108) &
    Office Products - Office \& School Supplies - Staplers \& Punches - Staples \\

    5 & Arrow Fastener 224 Genuine P22 1/4-Inch Staples, 5,050-Pack &
    Office Products - Office \& School Supplies - Staplers \& Punches - Staples \\

    \bottomrule
    \end{tabular}
    \caption{Example input sequence and recommendations from \flare with various alternative category-based critiques.}
    \label{tab:alt_traces}
\end{table*}

\end{document}